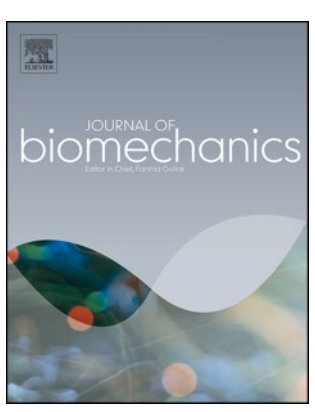

Review

# What the PCSA? Addressing diversity in lower-limb musculoskeletal models: age- and sex-related differences in PCSA and muscle mass

R. Maarleveld [a], H.E.J. Veeger [a], F.C.T. van der Helm [a], J. Son [b], R.L. Lieber [c,d,e,f], E. van der Kruk [a,*]

[a] Department of Biomechanical Engineering, Faculty of Mechanical Engineering, Delft University of Technology, the Netherlands
[b] Department of Biomedical Engineering, New Jersey Institute of Technology, Newark, NJ 07102, USA
[c] Shirley Ryan AbilityLab, Chicago, IL 60611, USA
[d] Department of Physiology, Northwestern University, Chicago, IL 60611, USA
[e] Department of Physical Medicine & Rehabilitation, Northwestern University, Chicago, IL 60611, USA
[f] Research Service, Hines VA Hospital, Maywood, IL 60153, USA

ABSTRACT

Musculoskeletal (MSK) models offer a non-invasive way to understand biomechanical loads on joints and tendons, which are difficult to measure directly. Variations in muscle strength, especially relative differences between muscles, significantly impact model outcomes. Typically, scaled generic MSK models use maximum isometric forces that are not adjusted for different demographics, raising concerns about their accuracy. This review provides an overview on experimentally derived strength parameters, including physiological cross-sectional area (PCSA), muscle mass (Mm), and relative muscle mass (%Mm), which is the relative distribution of muscle mass across the leg. Limited lower extremity PCSA data prevented assessment of differences in PCSA distribution. We analysed differences by age and sex, and compared open-source lower limb MSK model parameters with experimental data from 57 studies. Our dataset, with records dating back to 1884, shows that uniformly increasing all maximum isometric forces in MSK models does not capture key age-and sex-related differences in muscle ratio. Males have a significantly higher proportion of muscle mass in the rectus femoris (12%) and semimembranosus(15%) muscles, while females have a greater relative muscle mass in the pelvic (gluteus maximus(17%) and medius(23%)) and ankle muscles (tibialis anterior(14%) and posterior(15%), and extensor digitorum longus(16%)). Older adults have a higher relative muscle mass in the gluteus medius(37%), while younger individuals show more in the gastrocnemius(31%). Current MSK models do not accurately represent muscle mass distribution for specific age or sex groups. None of them accurately reflect female muscle mass distribution. Further research is needed to explore musculotendon age- and sex differences.

## 1. Introduction

Over the past three decades, musculoskeletal (MSK) computer models have served as a tool for understanding biomechanical loads on joints and tendons during movements that are difficult or impossible to measure experimentally. These models are valuable for studying musculoskeletal pathologies, optimizing movement in sports, assessing surgical interventions, and developing assistive devices (Delp et al., 2007; Grabke & Andrysek, 2018; Van Der Kruk & Geijtenbeek, 2024).

The most commonly used MSK models are defined by rigid bodies, joints, and muscles, representing the musculoskeletal system in a generalized way (Delp et al., 2007; Rajagopal et al., 2016). However, muscle parameters such as muscle path, muscle moment arm, tendon slack length, optimal fiber length, activation and deactivation dynamics, and maximum isometric force vary between individuals. To scale a generic musculoskeletal model, one or more of these parameters are adjusted to match the characteristics of a specific population or individual. Studies have shown that force predictions in MSK simulations are sensitive to the parameters of Hill-type muscle models, particularly optimal fiber length, physiological cross-sectional area (PCSA), and tendon slack length (Heinen et al., 2016; Redl et al., 2007; Scovil & Ronsky, 2006).

* Corresponding author.
*E-mail address:* e.vanderkruk@tudelft.nl (E. van der Kruk).

Most biomechanical studies rely on scaled open-source generic models, such as those in OpenSim (Delp et al., 2007; Seth et al., 2018), which are based on a combination of a limited number of post mortem human tissue (ex-vivo, in-vitro) studies and some in-vivo data. Except for one lower extremity model (Horsman, 2007; Klein Horsman et al., 2007; Modenese et al., 2011), these generic models (Arnold et al., 2010; Delp et al., 1990; Rajagopal et al., 2016) are based on a combination of data from different sources using mixed-sex data to represent muscle parameters (Anderson & Pandy, 1999; Friederich & Brand, 1990; Handsfield et al., 2014; Ward et al., 2009; Wickiewicz et al., 1983; Yamaguchi & Zajac, 1989), and bone geometry derived from male cadavers (Delp et al., 1990). In this process, segment properties are scaled linearly to an individual or population, with the parameters optimal fiber length and tendon slack length scaled according to segment length. However, PCSA, and thus maximum isometric force parameters, are often left unscaled (OpenSim Confluence, 2024b). Researchers typically increase these values uniformly when simulations fail, but no standardized guidelines exist for appropriately scaling. While it is widely recognized that aging affects the body and that individuals differ in body shape and composition, these factors are typically not incorporated into the maximum isometric strength parameters of the models.

Moreover, incorporating age-differences also requires understanding of the relationship between age and sex. For example, research showed that isokinetic strength in older men is lower than in young men but comparable to that of young women even when corrected for body mass (van der Kruk et al., 2022). Musculoskeletal disorders often present differently between sexes and throughout the lifespan. Osteoarthritis is more common in older women, while men experience higher rates of herniated discs(O'Connor, 2007). Young women face greater risks of ACL tears and patellar dislocation, whereas men are more prone to hamstring strains (Dai et al., 2024; Lin et al., 2018; O'Sullivan & Tanaka, n.d.). Ignoring age-sex differences therefore compromises the power and applicability of musculoskeletal models to improve our understanding and treatment of musculoskeletal health.

### 1.1. Maximum isometric force

In musculoskeletal modeling, maximum isometric force ($F_m^0$) is the highest contractile force produced by a fully activated muscle at its optimal fiber length and zero contraction velocity. This force depends on the number of parallel sarcomeres, represented by the physiological cross-sectional area (PCSA), and the muscle-specific tension ($\sigma_{muscle}$), which is the ability of the sarcomeres to produce force:

$$F_m^0 = PCSA \bullet \sigma_{muscle} \tag{1}$$

For muscles with fibers running parallel to the long axis, such as the hamstrings, PCSA is equal to the anatomical cross-sectional area (ACSA) and can be measured at the thickest part of the muscle. In pennate muscles, such as the rectus femoris, PCSA is measured perpendicular to the fibers, resulting in a V- or arc-shaped cut (Fick, 1910) (Fig. 1a). Pennate muscles generate greater strength than muscles with parallel fibers and similar ACSA due to their larger PCSA (i.e., larger number of fibers).

### 1.2. Relative distribution of maximum isometric forces

The ratio of maximum isometric forces, which is the relative distribution of maximum isometric forces across muscles, is a key factor because of the redundancy solver used in opensource MSK simulation software (Geijtenbeek, 2019; Seth et al., 2018). This mathematical solver resolves muscle redundancy by minimizing a cost function, often reducing muscle activation (Geijtenbeek, 2019; Seth et al., 2018) and thereby favouring muscles with higher maximum isometric forces (Supplementary Material A.1). Therefore, optimizations are sensitive to the relative differences of maximum isometric force across muscles, raising concerns about the validity of using the same ratios for different populations.

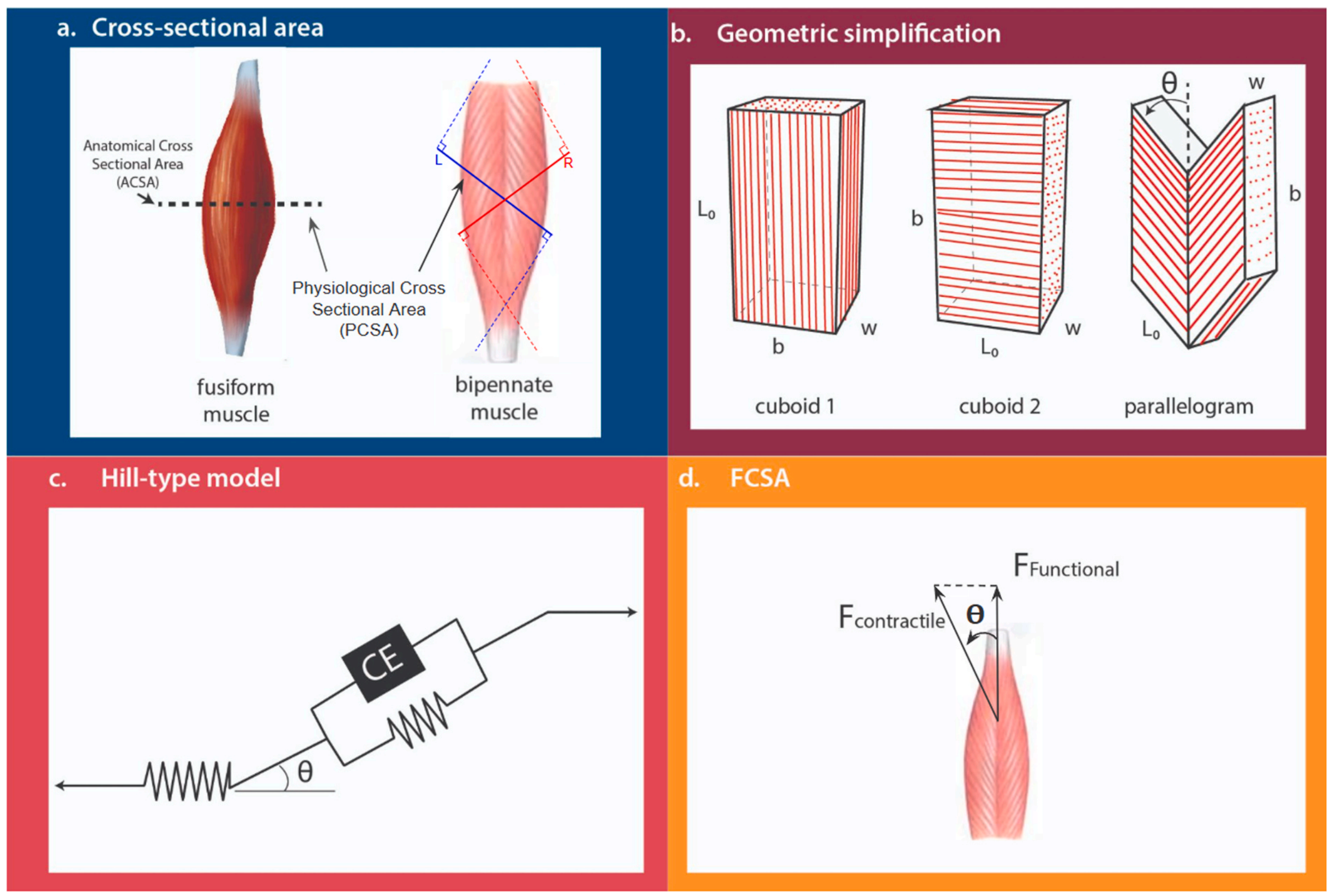

**Fig. 1.** a) ACSA is the anatomical cross sectional area, PCSA is the cross sectional area perpendicular to the muscle fibers; b) geometric simplification used to estimate PCSA from muscle volume. cuboid 1 represents a muscle with the muscle fibers in line with the muscle tendon, cuboid 2 represents a theoretical situation in which the muscle fibers have a pennation angle of 90 degrees,the parallelogram represents a bipennate muscle. The dotted area is the PCSA C) Hill type muscle model with an active and a passive element in parallel and a passive element in series with pennation angle ($\theta$). This representation does not reflect the geometrical behavior of the muscle, but should be interpreted as a 1D model representing the muscle function. D) projection of the muscle force in the direction of the tendon ("functional" force).

Although differences in specific tension among fiber types and motor units have been observed in animal studies (S. Bodine et al., 1987; S. C. Bodine et al., 1988), most musculoskeletal models assume that $\sigma_{muscle}$ is constant across muscles and individuals. This means that $\sigma_{muscle}$ acts as a scaling factor for maximum isometric force, with PCSA being the main determinant of the distribution between muscles.

Despite the increasing use of medical imaging to estimate PCSA data, there are no reference materials for scaling muscle parameters across specific age or sex groups. Furthermore, it remains unclear which of these groups are adequately represented by the most frequently downloaded open-source MSK models. This raises the question of whether current practices in scaling MSK models effectively account for differences in PCSA between sex and age groups.

### 1.3. Aim

The goal of this *meta*-analysis was to use the literature to understand how lower limb muscle PCSA, and thus, maximum isometric force, should be scaled relative to sex and age. This review aims to provide a comprehensive overview of experimentally derived PCSA parameters, focusing on PCSA as an indicator of maximum isometric strength. This approach avoids bias introduced by variations in muscle-specific tension ($\sigma_{muscle}$) when comparing maximum isometric strength between literature values and those used in OpenSim models. The study differentiates among various age groups, sexes, and fitness levels, examines distribution of PCSA (%PCSA), and compares these findings to the muscle parameters used in generic open-source lower limb MSK models.

### 1.4. PCSA definition

Although direct measurement of PCSA is possible (Veeger et al., 1991), it is challenging, especially for pennate muscles. Therefore, proxies have frequently been used. These approximations simplify muscle geometry, assuming it as a cuboid, and calculate cross-sectional areas based on this approximation (Fig. 1b). The PCSA is directly proportional to muscle volume and inversely proportional to fiber length (Alexander, 1975; Fick, 1910; Pfuhl, 1937), where nowadays the optimal fiber length is used (Lieber & Fridén, 2000):

**Method 1:**

$$PCSA = \frac{M_m}{L_o \bullet \rho} \tag{2}$$

Where $M_m$ is the muscle mass, $L_o$ is optimal fiber length, and $\rho$ is the muscle density which in the models is kept contact at 1.056 g/cm$^3$ (Méndez, 1960). Dissection studies used muscle mass as it was easier to determine than volume. Modern imaging simplifies this, allowing PCSA to be calculated as

**Method 2:**

$$PCSA = \frac{V_m}{L_o} \tag{3}$$

where $V_m$ is the muscle volume. This method is applicable in vivo at specific muscle lengths (joint angles) and tension levels.

### 1.5. Pennation angle

The magnitude of the force that a muscle can produce is unrelated to the pennation angle ($\theta$). However, $\theta$ does influence the direction of the force and is needed to determine the force in the longitudinal direction, working on the tendon using the projection of PCSA (Fick, 1910). In line with previous publications, we will refer to this as the *functional* cross-sectional area (FCSA) (Powell et al., 1984; Rockenfeller et al., 2024):

**Method 1b:**

$$FCSA = \frac{M_m}{L_o \bullet \rho} \bullet cos(\theta) \tag{4}$$

**Method 2b:**

$$FCSA = \frac{V_m}{L_o} \bullet cos(\theta) \tag{5}$$

Correcting for the pennation angle isolates the force acting on the tendon, excluding lateral forces affecting muscle thickening and thinning. This correction is valuable in some applications, such as for determining effective strength between muscles (Brand et al., 1981).

In musculoskeletal modeling, maximum isometric force ($F_m^0$) is a key parameter in the contractile element of Hill's muscle model, which includes a contractile element in series with an elastic spring (tendon) and a parallel elastic element (Fig. 1c). This model incorporates force–length and force–velocity relationship curves, adjusted by $F_m^0$. The pennation angle (θ), though not part of the original model, converts the parallel elements into a force aligned with the tendon via cos(θ) (Fig. 1c).

As the model inherently accounts for this force projection, *FCSA* should not be used for $F_m^0$ in musculoskeletal simulations. Instead, $F_m^0$ should be calculated using PCSA as outlined in Method 1 or 2. Since the 1980s, many studies have reported *FCSA* (Barber et al., 2011a; Barber et al., 2011b; Blazevich et al., 2009; Charles et al., 2019; Fukunaga et al., 1992, 1996; Narici et al., 1992; Son et al., 2024; Ward et al., 2009; Wickiewicz et al., 1983). Using these data directly for $F_m^0$ leads to an underestimation by a factor of 1/cos(θ). This error varies between muscles due to differences in pennation angles.

Separate from this discrepancy, is the ongoing debate about using the pennation angle in Hill-type models (Lieber, 2022). Static pennation angles do not accurately reflect dynamic muscle behavior during contraction, and even dynamic models fail to account for shear forces between muscle fibers.

In this review, we have included only those articles that reported the PCSA according to Method 1 and 2, or could be corrected to represent these methods. Each approach to calculating PCSA requires specific muscle parameters. Initially, these were assessed through cadaver dissections (Supplementary material A.2), which continue today (Son et al., 2024). Recently, newer datasets have emerged using advanced imaging techniques like MRI, US, and DTI. Supplementary material A.3) outlines current methods for measuring PCSA parameters.

## 2. Method

### 2.1. Search strategy

A literature search was performed in the Scopus, Pubmed, and Google Scholar databases between April 2023 to July 2024. The following keywords were used: *("Physiological cross-sectional area" OR "PCSA" OR "Muscle Volume") AND ("Pelvis" OR "Pelvic" OR "Shank" OR "Thigh" OR "Leg" OR "Foot" OR "extensors" OR "Flexors" OR "Lower limb" OR "-specific muscle name-") AND ("MRI" OR "MR Imaging" OR "Magnetic resonance imaging" OR "CT" OR "computer tomography" OR "Ultrasound" OR "cadaver" OR "Dissection" OR "in-vivo" OR "ex-vivo").*

The bibliography of the located studies was thoroughly reviewed to ensure that all relevant works were included, even those that were inadvertently omitted from the keyword-based search.

The studies were initially selected based on the relevance of their titles and abstracts. Subsequently, studies that provided explicit values for PCSA for healthy individuals were included. Non-human studies and numerical simulation studies were excluded. Articles presenting only graphical representations of PCSA outcomes without numerical data were also excluded. Additionally, intrinsic foot muscles were omitted from the analysis since they are often not included in current lower limb MSK models. When only summarized data were presented, original data were requested from the authors.

## 2.2. Data extraction

### 2.2.1. Data from measurements

Papers were organized based on relevant topics, including the lower limb location (leg, pelvis), participant group (children, young adults, adults, elderly, athletes), muscle-specific tension, and image modality reliability. We categorized the reported values into four age groups according to the average onset of muscle mass decline at 30 years and the onset of accelerated muscle mass decline at 65 years (van der Kruk et al., 2021):

- Children (CH): 18 years and younger
- Young Adults (YA): 18–30 years
- Adults (AD): 30–65 years
- Elderly (EL): 65 years and older

If a study explicitly mentioned that reported values were from athletes, we categorized them under the label ’Ath’, followed by the respective age group. We utilized two sex groups (female (F) and male (M)) to determine the ratio and number of individuals of each sex within each category.

We gathered the following additional information from each paper:

- Imaging modality
- Mean, Standard deviation (SD), and range of PCSA values.
- Sample size
- Male-Female ratio
- Participant characteristics (weight, height, age)
- Body position during measurements
- Muscle characteristics: (optimal) fiber length & pennation angle
- Method for determining PCSA
- Specific strength

### 2.2.2. Data from generic open-source MSK models

We included the model parameters of the five most downloaded OpenSim lower limb models(Simtk, 2023) in the comparison Table 1.

- **Delp (Model D)** (Delp et al., 1990), simulates a 1.8-meter tall individual with a bodymass of 75.16 kg using muscle parameters primarily sourced from Wickiewicz et al. (1983), supplemented by data from Friederich & Brand (1990), with the lower leg represented by 13 rigid-body segments interconnected through various joints, resulting in 14 degrees of freedom. The model has 43 muscle–tendon actuators.
- **Gait2392/2354 (Model G)** (OpenSim Confluence, 2024a), is a three-dimensional representation of a 1.80-meter tall individual weighing 75.16 kg, incorporates muscle parameters from Delp but scaled to align with joint torque–angle relationships measured in living subjects (Carhart, 2000). The model’s lower extremity feature 13 rigid-body segments and 23 degrees of freedom with joint definitions stem from Delp, while low back joint and anthropometry details are sourced from Anderson & Pandy (1999), with a planar knee model from Yamaguchi & Zajac (1989); The model has 92 (Gait2392) and 54 (Gait2354) muscle–tendon actuators.
- **Arnold (Model A)** (Arnold et al., 2010), depicting a 1.70 m tall individual, integrates model geometry from Delp but adopts muscle architecture detailed by Ward et al. (2009), featuring 14 rigid-body segments for the lower leg, encompassing 23 degrees of freedom and incorporating muscle lines of action for 44 muscle–tendon actuators.
- **London Lower Limb (Model L)** (Horsman, 2007; Modenese et al., 2011), also known as the ‘London Lower Limb model’ (OpenSim., n. d.), depicts a human, utilizing muscle architecture detailed by Klein Horsman et al. (2007), with the lower leg modeled by 11 rigid-body segments and 12 degrees of freedom, and incorporating muscle lines of action for 163 actuators representing 38 muscles.
- **Rajagopal (Model R)** (Rajagopal et al., 2016), depicts a 1.7 m tall individual weighing 75 kg, utilizing model geometry from Arnold, and incorporates musculotendon parameters derived from anatomical measurements of cadaver specimens (Ward et al. (2009) and magnetic resonance images of young healthy subjects (Handsfield et al. (2014), featuring 13 rigid-body segments for the lower leg with 37 degrees of freedom and including muscle lines of action for 80 actuators in the lower limb and 17 ideal torque actuators driving the upper body.

**Table 1**
OpenSource Musculoskeletal models and their sources for muscle parameters.

| Model | Model description | | | | Dataset | | | | | |
|---|---|---|---|---|---|---|---|---|---|---|
| **Name** | **Segments** | **Actuators** | DOF | **Height (m)** | **Bodymass (kg)** | **Source Muscle Parameters** | **Samples (M:F)** | **Age (Years)** | **Height (m)** | **Bodymass (kg)** |
| Delp (Delp et al., 1990) | 13 | 43 | 14 | 1.8 | 75.2 | (Wickiewicz et al., 1983) | 3 (?) | | | |
| | | | | | | (Friederich and Brand, 1990) | 2(1:1) | 37 & 63 | 1.83 & 1.68 | 91 & 59 |
| Gait2392/2354 (OpenSim Confluence, 2024a) | 13 | 92/ 54 | 23 | 1.8 | 75.2 | (Delp et al., 1990) | | | | |
| Arnold (Arnold et al., 2010) | 14 | 44 | 23 | 1.7 | | (Ward et al., 2009) | 21(9:12) | 83 ± 9 | 1.68 ± 0.09 | 82.7 ± 15.3 |
| Londen Lower Limb (Horsman, 2007; Modenese et al., 2011) | 11 | 163 | 12 | | | (Klein Horsman et al., 2007) | 1 (1:0) | 77 | 1.74 | 105 |
| Rajagopal (Rajagopal et al., 2016) | 13 | 97 | 37 | 1.7 | 75 | (Ward et al., 2009) | 21 (9:12) | 83 ± 9 | 1.68 ± 0.09 | 82.7 ± 15.3 |
| | | | | | | (Handsfield et al., 2014) | 24 (16:8) | 25.5 ± 11.1 | 1.71 ± 0.01 | 71.8 ± 14.6 |

### 2.3. Data analysis

#### 2.3.1. Relative muscle strength (%Mm)

To investigate age- and sex-related differences in %PCSA, we identified studies that reported PCSA data for individual males and females, focusing on those that included data for the muscles of at least one complete leg. Due to the limited availability of such studies in young adults, we chose to use relative muscle mass (%Mm) instead of %PCSA, as it allowed us to draw from a broader dataset. In doing so, we assumed, consistent with the opensource musculoskeletal models, that muscle density is constant across muscles within a healthy individual. It is important to note that optimal fiber length lies between muscle mass and PCSA (eq.2–7). In theory, if age-and sex related relative differences in optimal fiber length were perfectly inversely proportional to corresponding differences in muscle mass, this could eliminate observed differences in %PCSA. However, we consider such a scenario unlikely. As a result, differences in %Mm still provide a meaningful and relevant approximation.

We estimated %Mm by dividing the mass of each individual muscle by a minimal set of lower limb muscles (Charles et al., 2019): adductor brevis (AB), adductor longus (AL), adductor magnus (AM), gracilis (GRA), biceps femoris (BIC), semimembranosus (SM), semitendinosus (ST), rectus femoris (RF), vastus lateralis (VL), vastus medialis (VM), tibialis anterior (TA), extensor digitorum longus (EDL), extensor hallucis longus (EHL), soleus (SOL), gastrocnemius (GAS), and sartorius (SAR).

To enable comparison among studies, we combined the masses of related muscles to form muscle groups: gastrocnemius (GAS = medial gastrocnemius (MG) + lateral gastrocnemius (LG)), vastus (VAS = vastus intermedius (VI) + vastus lateralis (VL) + vastus medialis (VM)), iliopsoas (ILIOPS = iliacus (Iliac) + psoas major (Psoas)), and biceps femoris (BIC = biceps femoris short head (BFSH) + biceps femoris long head (BFLH)).

We categorized the data into defined sex (F,M) and age (YA, AD, EL) groups and conducted a two-way ANOVA to detect significant differences in %Mm.

#### 2.3.2. Absolute muscle strength (PCSA and Mm)

To provide a comprehensive overview of all PCSA values, including those that did not report on a full lower leg, we presented PCSA values in a graphical format, categorized by age, sex, and fitness level. The graphic depicted the mean PCSA values, and if available, the range. In cases where the range was not provided, the 95 % confidence interval (CI) was estimated using the formula:

$$CI_{95} = PCSA_{mean} \pm Z \bullet \frac{PCSA_{SD}}{\sqrt{n}} \tag{8}$$

In which Z = 1.96 for n ≥ 30 and the t-distribution value for Z when n < 30.

For the articles used to estimate %Mm, we also reported the absolute muscle mass (Mm) and categorized the data into defined age and sex groups. We conducted a two-way ANOVA to detect significant differences in Mm.

#### 2.3.3. Comparison generic open-source MSK models

In each of the graphical overviews, we integrated %Mm, Mm, and PCSA values from the generic open-source MSK model data. For each muscle, we evaluated the model's representation across specific age-sex-fitness groups. If a particular group had a sample size of fewer than five individuals for a muscle parameter measurement, we deemed the data insufficient for drawing conclusions.

To enable comparison of %Mm with the generic models, we determined the muscle masses in the models using the formula:

$$M_{m-model} = PCSA \bullet L_o \bullet \rho \tag{9}$$

In which $L_o$ is optimal fiber length, and $\rho$ is the muscle density (1.056 g/cm3 (Méndez, 1960)). We then obtained %Mm by dividing the mass of each individual muscle by the total mass of the previously mentioned minimal set of lower limb muscles.

## 3. Results

### 3.1. Search results

Initially, 879 papers were identified, with 330 remaining after pre-selection. After removing duplicates, 322 papers were reviewed, and 57 containing explicit PCSA data were included for further analysis. The flow diagram of the systematic review is provided in Fig. 2.

Table 2 shows the number of participants per demographic group for each muscle. A graphical overview presenting all PCSA values are provided in Supplementary Tables A.4.2-A.4.4 and the optimal fiber lengths used in each study is found in Supplementary Tables A.4.5-A.4.8 Most data are available for the knee flexors (SM, ST) and extensors (RF, VAS), and larger plantar flexors (GAS), primarily for young adults and the elderly. However, males are underrepresented in the elderly group, and females in the young adult group. There is also a lack of data for children (<18 years) and adults (30–65 years) (Table 2).

To determine %Mm, we identified articles that reported muscle mass (Mm) for at least one full leg and separately for males and females (Supplementary Material A.2): (Charles et al., 2019; Friederich & Brand, 1990; Klein Horsman et al., 2007; Ruggiero et al., 2016; Son et al., 2024; Theile, 1884; Ward et al., 2009). Studies lacking participant characteristics were excluded. Seireg & Arvikar (1973) and Wickiewicz et al. (1983) (Fig. 2). %Mm was chosen over %PCSA due to limitations in using these articles to estimate PCSA, as some studies did not measure optimal fiber length (Theile, 1884) or based their calculations on mean fiber length (Friederich & Brand (1990), (Charles et al., 2019)). As a result, no %PCSA data were available for young individuals. Fig. 3 offers a visual overview of the main findings discussed in the next sections.

### 3.2. Sex-differences

#### 3.2.1. Relative and absolute muscle mass

Males had significantly higher Mm in muscles compared to females, except for LG, Gmin, and Piri, where the differences were not statistically significant (Fig. 4). In one muscle, MG, the Mm age groups differed significantly in mean age (females: 71 years, males: 53 years).

While males had higher absolute Mm, these differences mostly disappeared when considering %Mm (Fig. 5). However, males had significantly higher %Mm in RF ($p = 0.028$) and SM ($p = 0.026$), indicating a greater proportion of muscle mass in these biarticular muscles compared to females (RF = 5.5 %, SM = 6 % in males vs. RF = 4.9 %, SM = 5.2 % in females). Conversely, females showed higher %Mm in pelvic muscles gluteus maximus (Gmax: F = 26 %, M = 21.7 %, $p = 0.031$) and gluteus (Gmed: F = 12.6 %, M = 9.7 %, $p = 0.015$), as well as in ankle muscles tibialis anterior (TA: F = 3.7 %, M = 3.2 %, $p = 0.006$), tibialis posterior (TP: F = 2.7 %, M = 2.3 %, $p = 0.038$), and extensor digitorum longus (EDL: F = 1.9 %, M = 1.6 %, $p = 0.032$). Piri and QdF were also higher in female, but small sample sizes limit conclusions. There were no significant %Mm differences across age groups.

#### 3.2.2. Comparison generic open-source MSK models

Comparing Mm in the musculoskeletal models with experimental data, the Rajagopal and Londen Lower Limb models showed the most consistent representation, generally exceeding male Mm means and best representing male absolute muscle mass, with a few exceptions. The Arnold model mostly fell between male and female means, suggesting it could represent a stronger than average female or a weaker than average male. Both the Gait2392/2354 and Delp model fluctuated widely, with some muscles exceeding male means and others falling in the low female range, making them inconsistent.

The Gait2392/2354 and Delp models fell outside experimental ranges for multiple muscles, inconsistently representing muscle mass distribution for either sex. The Arnold, Londen Lower Limb, and Rajagopal models were within experimental ranges. For muscles with significant sex differences in %Mm, Londen Lower Limb and Rajagopal matched male distribution. Arnold aligned with the female distribution for RF, EDL, and Gmed, and the male distribution for TA, TP, and Gmax.

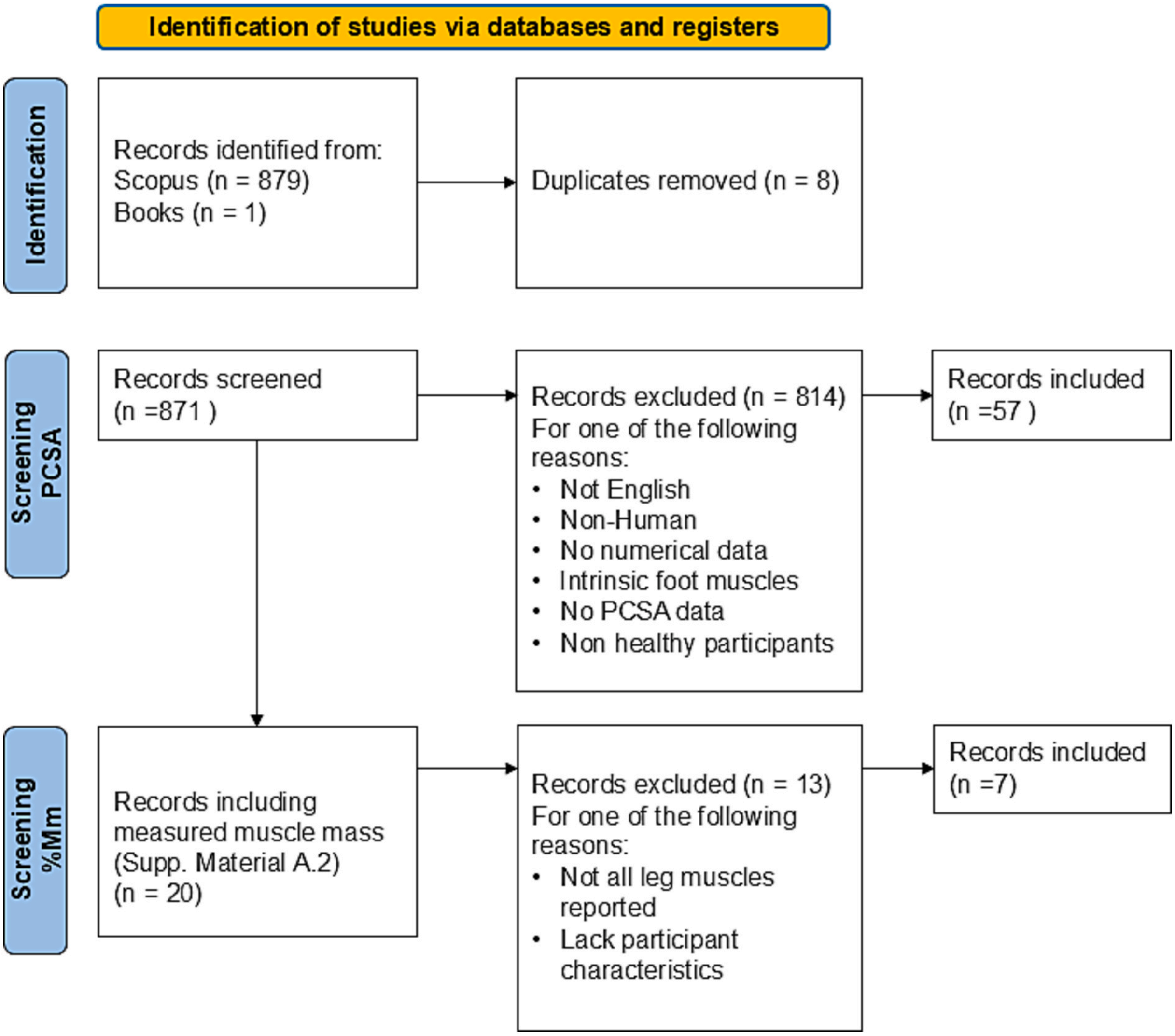

**Fig. 2.** Flow diagram of the systematic review, identifying papers for the PCSA analysis and the Mm analysis.

### *3.3. Age-differences*

#### *3.3.1. Relative and absolute muscle mass*

There were insufficient experimental data for Pect, QdF, Piri, TFL, and Gmin in the elderly group (n = 2) for age-group comparison, and for BFSH, BFLH, LG, Ps, and FDL in the young group (n < 5), but enough for the combined BIC and GAS. In muscles where data were available, the elderly had significantly less Mm than the young, except for ST ($p = 0.072$), VL ($p = 0.076$), VI ($p = 0.087$), PerB ($p = 0.066$), Gmax ($p = 0.156$), and Gmed ($p = 0.987$). These non-significant differences still showed a trend of a lower mean Mm in the elderly, except for Gmed, where the means were similar. Adults also showed significantly lower Mm than the young in 14 muscles (Fig. 6). Only FHL had a significantly lower mean in the elderly compared to adults.

Most age differences disappeared in the %Mm comparison (Fig. 7, Fig. 3), excluding muscles with fewer than five samples per group. Young individuals had significantly higher %Mm compared to adults and the elderly for RF (YA = 6.1 %, AD = 5.2 %, EL = 5 %, $p = 0.007$), GAS (YA = 10.2 %, AD = 8.5 %, EL = 7.8 %, $p < 0.001$), and MG (YA = 7.9 %, AD = 5.5 %, EL = 4.9 %, $p = 0.001$). A significant age × sex interaction was found for GAS ($p = 0.005$), with young individuals having higher %Mm compared to the elderly for FHL (YA = 2.5 %, EL = 1.7 %, $p = 0.042$). Gmed was significantly *higher* in the elderly than in the young (YA = 7.6 %, EL = 12 %, $p = 0.042$). However, the data for young individuals, especially females (Gmed: F/M = 1/4), were limited. A main effect of age was observed for SM, but post hoc analysis did not show significant differences between groups.

#### *3.3.2. Comparison generic open-source MSK models*

Arnold is the most suitable for representing the elderly, with both Mm and %Mm closely matching the elderly group mean across all muscles. Rajagopal best reflects young adults' Mm, except for ST, which aligns more with elderly values. For %Mm, Rajagopal is mostly consistent with adults across several muscles, making it the best option for non-elderly populations. Londen Lower Limb generally exhibits higher Mm, making it representative of young adults. However, for %Mm, it better reflects the elderly, particularly for muscles like Gmed, FHL, and GAS, which have significant age-related differences. Gait2392/2354 is inconsistent, with Mm and %Mm fluctuating between young and elderly across muscles, making it unrepresentative of any specific age group. Delp aligns with the elderly for Mm but fluctuates in %Mm across age

**Table 2**
Number of participants per demographic group for each muscle in the PCSA analysis of 57 studies. The full specifications for each of the muscles can be found in Supplementary Material A.4. In this table CH = Children (<18 years), YA = Young Adults (18–30 years), A = Adults (30–65 years), and E = Elderly (>65 years).

| | *Muscle* | CH ♀ | CH ♂ | YA ♀ | YA ♂ | AD ♀ | AD ♂ | EL ♀ | EL ♂ |
|---|---|---|---|---|---|---|---|---|---|
| **Knee flexors** | *BFSH* | 0 | 0 | 8 | 16 | 7 | 8 | 14 | 4 |
| | *BFLH* | 0 | 0 | 8 | 16 | 6 | 8 | 14 | 4 |
| | *SM* | 0 | 0 | 11 | 19 | 6 | 7 | 28 | 18 |
| | *ST* | 0 | 0 | 11 | 19 | 6 | 8 | 14 | 4 |
| **Knee extensors** | *RF* | 0 | 0 | 11 | 80 | 7 | 8 | 11 | 1 |
| | *VL* | 9 | 21 | 16 | 226 | 17 | 16 | 40 | 21 |
| | *VM* | 0 | 0 | 11 | 80 | 7 | 8 | 11 | 1 |
| | *VI* | 0 | 0 | 11 | 80 | 7 | 8 | 11 | 1 |
| **Plantar flexors** | *SOL* | 0 | 0 | 8 | 39 | 7 | 8 | 11 | 9 |
| | *MG* | 12 | 18 | 97 | 100 | 33 | 8 | 107 | 45 |
| | *LG* | 0 | 11 | 8 | 67 | 7 | 8 | 11 | 39 |
| | *TP* | 0 | 0 | 8 | 16 | 7 | 8 | 11 | 9 |
| | *PerB* | 0 | 0 | 0 | 0 | 7 | 8 | 11 | 9 |
| | *PerL* | 0 | 0 | 0 | 0 | 7 | 8 | 0 | 9 |
| | *FDL* | 0 | 0 | 8 | 16 | 7 | 8 | 0 | 9 |
| | *FHL* | 0 | 0 | 8 | 16 | 7 | 8 | 0 | 9 |
| **Dorsi flexors** | *TA* | 0 | 0 | 8 | 16 | 7 | 8 | 11 | 9 |
| | *EDL* | 0 | 0 | 0 | 0 | 6 | 7 | 0 | 9 |
| | *EHL* | 0 | 0 | 0 | 0 | 7 | 8 | 0 | 9 |
| **Hip** | *Gmax* | 0 | 0 | 11 | 19 | 7 | 8 | 13 | 5 |
| | *Gmed* | 0 | 0 | 11 | 19 | 7 | 8 | 11 | 1 |
| | *Gmin* | 0 | 0 | 11 | 19 | 1 | 1 | 0 | 1 |
| | *ILIA* | 0 | 0 | 8 | 16 | 7 | 8 | 11 | 1 |
| | *Ps* | 0 | 0 | 8 | 16 | 7 | 8 | 0 | 1 |
| | *SAR* | 0 | 0 | 11 | 19 | 7 | 8 | 11 | 1 |
| | *Piri* | 0 | 0 | 11 | 19 | 1 | 1 | 0 | 1 |
| | *TFL* | 0 | 0 | 11 | 19 | 1 | 1 | 11 | 1 |
| **Adductors** | *AB* | 0 | 0 | 11 | 19 | 7 | 8 | 11 | 1 |
| | *AL* | 0 | 0 | 11 | 19 | 6 | 7 | 11 | 1 |
| | *AM* | 0 | 0 | 11 | 19 | 7 | 8 | 11 | 1 |
| | *GRA* | 0 | 0 | 11 | 19 | 7 | 8 | 11 | 1 |
| | *PECT* | 0 | 0 | 11 | 19 | 1 | 1 | 0 | 1 |
| | *QdF* | 0 | 0 | 11 | 19 | 0 | 1 | 0 | 1 |

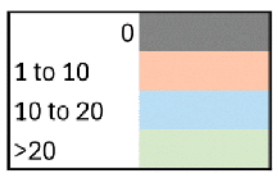

groups, with some muscles falling outside expected ranges.

### 3.4. Age-sex differences

#### 3.4.1. PCSA comparison generic open-source MSK models

The rules governing the scaling of optimal fiber length based on sex, age, and size remain unclear, and therefore there is no direct translation from Mm to PCSA from the experimental data (Son et al., 2024). Due to the limited availability of comprehensive lower limb PCSA data, we can only assess differences in PCSA, not %PCSA. However, the results indicate that conclusions derived from muscle mass (Mm) comparisons are also relevant to physiological cross-sectional area (PCSA) regarding the representation of age and sex in generic musculoskeletal (MSK) models (Supplementary Material A.4 Tables A.4.2-A.4.4).

## 4. Discussion

The purpose of this *meta*-analysis was to use the literature to understand how lower limb muscle PCSA, and thus, maximum isometric force, should be scaled relative to sex and age. Our results show that isometric scaling of maximum isometric force fails to account for age- and sex-related significant differences in relative muscle mass distribution. Current musculoskeletal models show varied representations for age and sex in absolute and relative distribution of muscle mass.

### 4.1. Sex-related differences

Males exhibited significantly higher absolute muscle mass compared to females, but these differences largely disappeared when examining the relative distribution of muscle mass across the leg, with a few exceptions (Fig. 3). Males showed significantly higher %Mm in RF and SM, indicating a greater proportion of muscle mass in these biarticular muscles. Conversely, females had much higher %Mm in pelvic muscles (Gmax and Gmed) and several ankle muscles (TA, TP, EDL), along with Piri and QF muscles, though limited samples for the latter two preclude definitive conclusions.

The increased %Mm in pelvic muscles among females may be attributed to anatomical differences in pelvic shape. Females typically have a wider pelvis, resulting in a larger insertion area for gluteal muscles, which could increase relative muscle mass in this region (Fischer & Mitteroecker, 2017). Additionally, differences in hip joint geometry between sexes lead to sex-related variations in moment arms (Cueto Fernandez et al., 2024), significantly correlated with Gmed volume (Preininger et al., 2011). In contrast, RF attaches to the iliac spine, which is positioned relatively further from the distal insertion in males due to their taller pelvic shape (Fischer & Mitteroecker, 2017). This results in a longer RF muscle in males compared to other lower limb muscles, which may lead to greater %Mm. Whether this holds true for %PCSA depends on whether optimal fiber length scales at the same linear rate with muscle length. A preliminary analysis using the limited available %PCSA data (Supplementary Material A.5) suggests that the sex difference remains present in %PCSA.

### 4.2. Age-related differences

The elderly exhibited significantly less absolute muscle mass compared to young individuals for most muscles, with exceptions ST, VL, VI, PerB, and Gmax, which showed non-significant differences but lower means in the elderly. Gmed mass appeared unaffected by age, consistent with findings by Preininger et al. (2011), which indicated no correlation between age and Gmed volume in 102 older patients. Most age-related differences diminished in %Mm comparisons, except for a few muscles. The young had significantly higher relative muscle mass in RF, MG, and GAS compared to adults and the elderly, and higher relative mass for FHL compared to the elderly. Notably, the elderly showed significantly higher relative muscle mass in Gmed compared to the young. Gmed's primary role in maintaining balance and stabilizing the pelvis contrasts with GAS and RF, which are more involved in dynamic activities like gait. This redistribution of muscle mass may reflect age-related changes in behavior and/or lifestyle. This observation aligns with reported findings in older adults. In daily life activities, there is a proximal redistribution of joint work, possibly linked to an uneven decline in muscle mass (Amiridis et al., 2003; Buddhadev & Martin, 2016; Horak, 2006; McGibbon & Krebs, 1999; Miller et al., 2024).

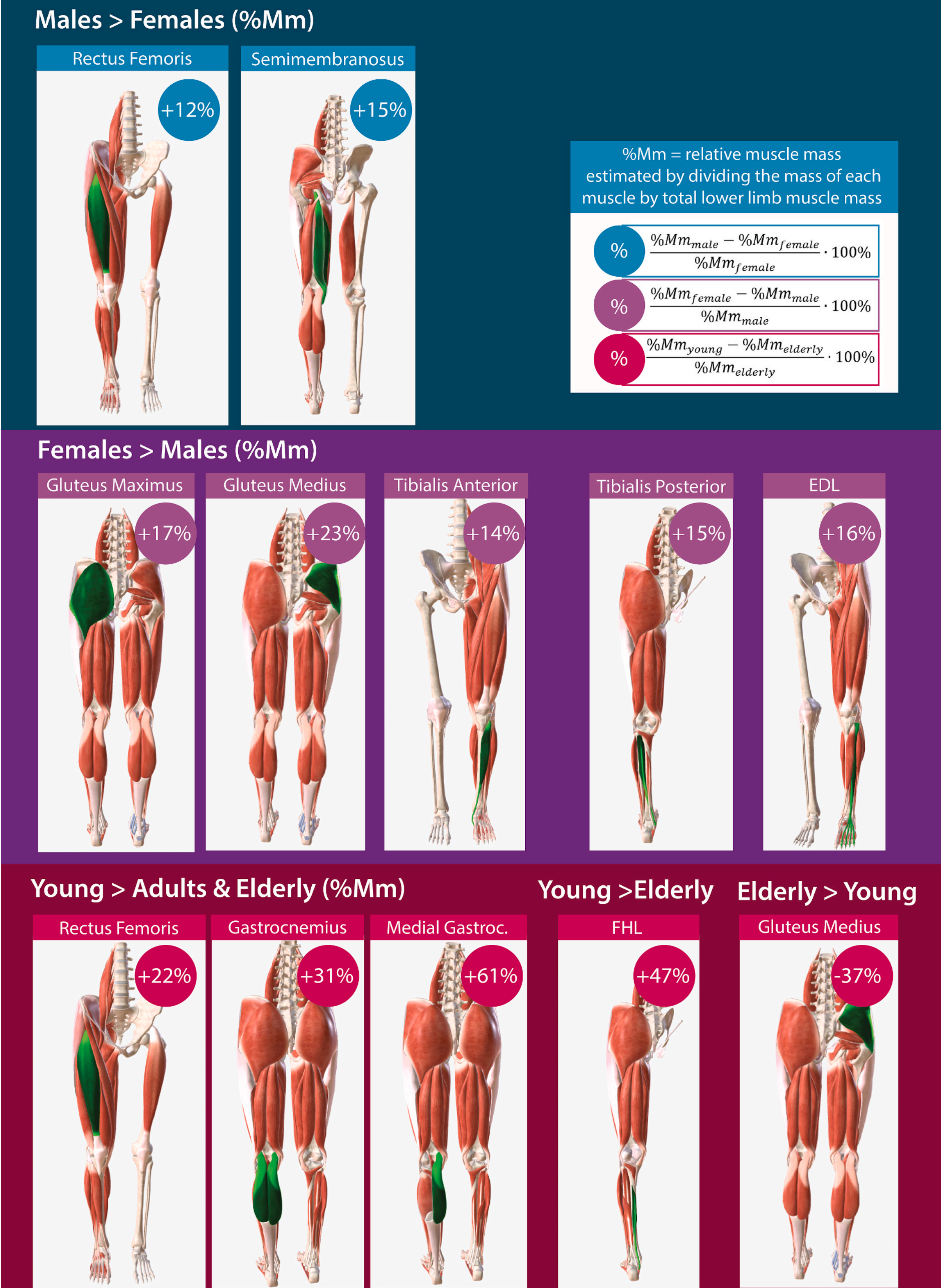

**Fig. 3.** Graphic summary of the muscles that had significant differences in %Mm between male and female, and young, adults, and elderly. In the circles, we indicated the percentual differences between the %Mm of groups. Females also had higher %Mm for the Piri and QF muscles compared to men, but due to the limited samples for these muscles we did not include them in this overview.

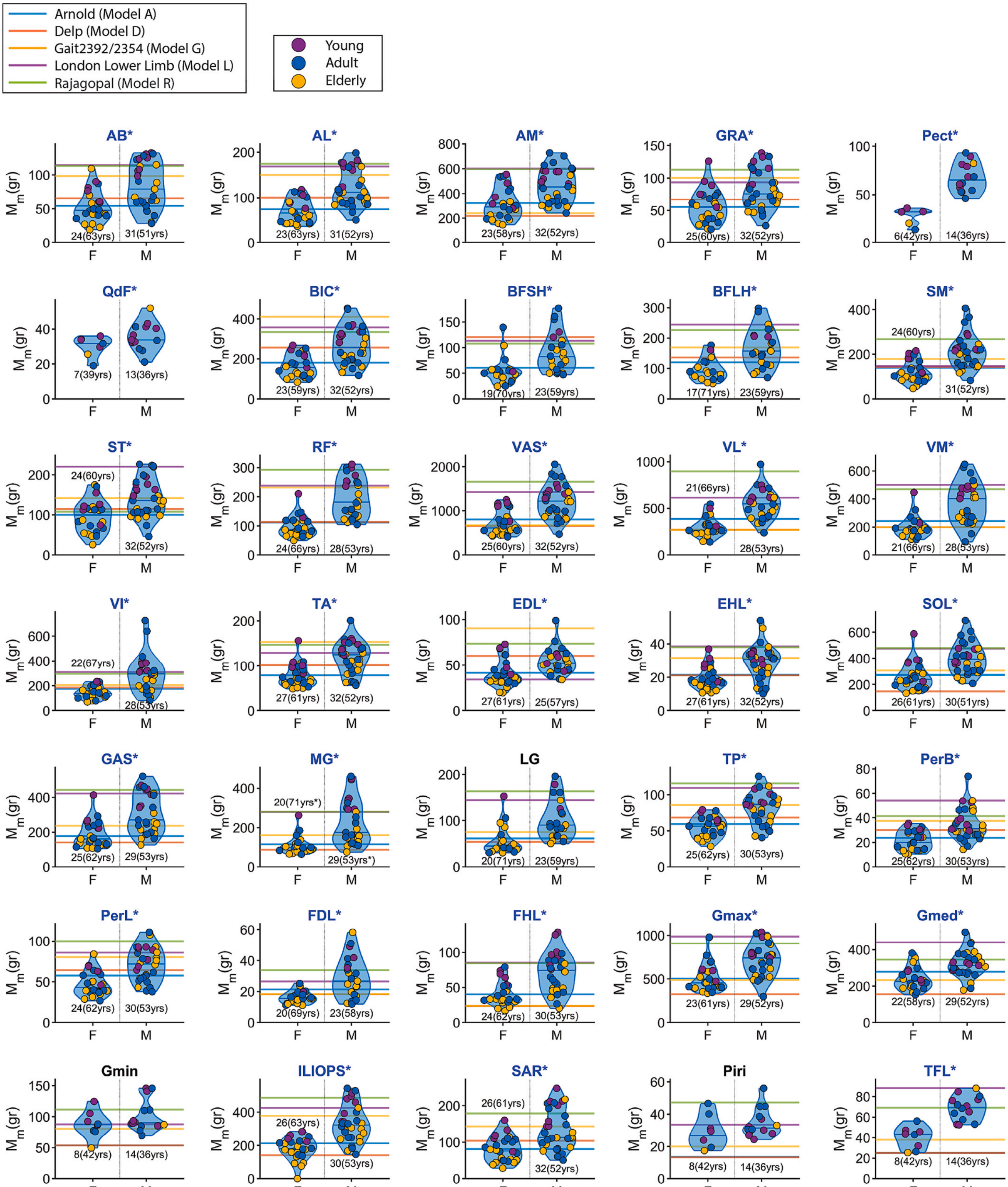

**Fig. 4.** Sex-related differences in absolute muscles mass (Mm) between females (F) and males (M). The colours in the dots indicate the age-groups (Young, Adult, Elderly). The title is coloured blue with an asterisk if a significant difference was found ($p < 0.05$). The horizontal lines indicate the muscle mass levels of the OpenSim opensource models. Some muscles were not incorporated in the musculoskeletal models (e.g. Pect, QdF). The number below the violin plots indicate the number of specimens (ex-vivo) included, this differs between muscles depending on the available data from the sources. In brackets is the average age, an asterisk there indicates that there is a significant difference in age between the groups. (For interpretation of the references to colour in this figure legend, the reader is referred to the web version of this article.)

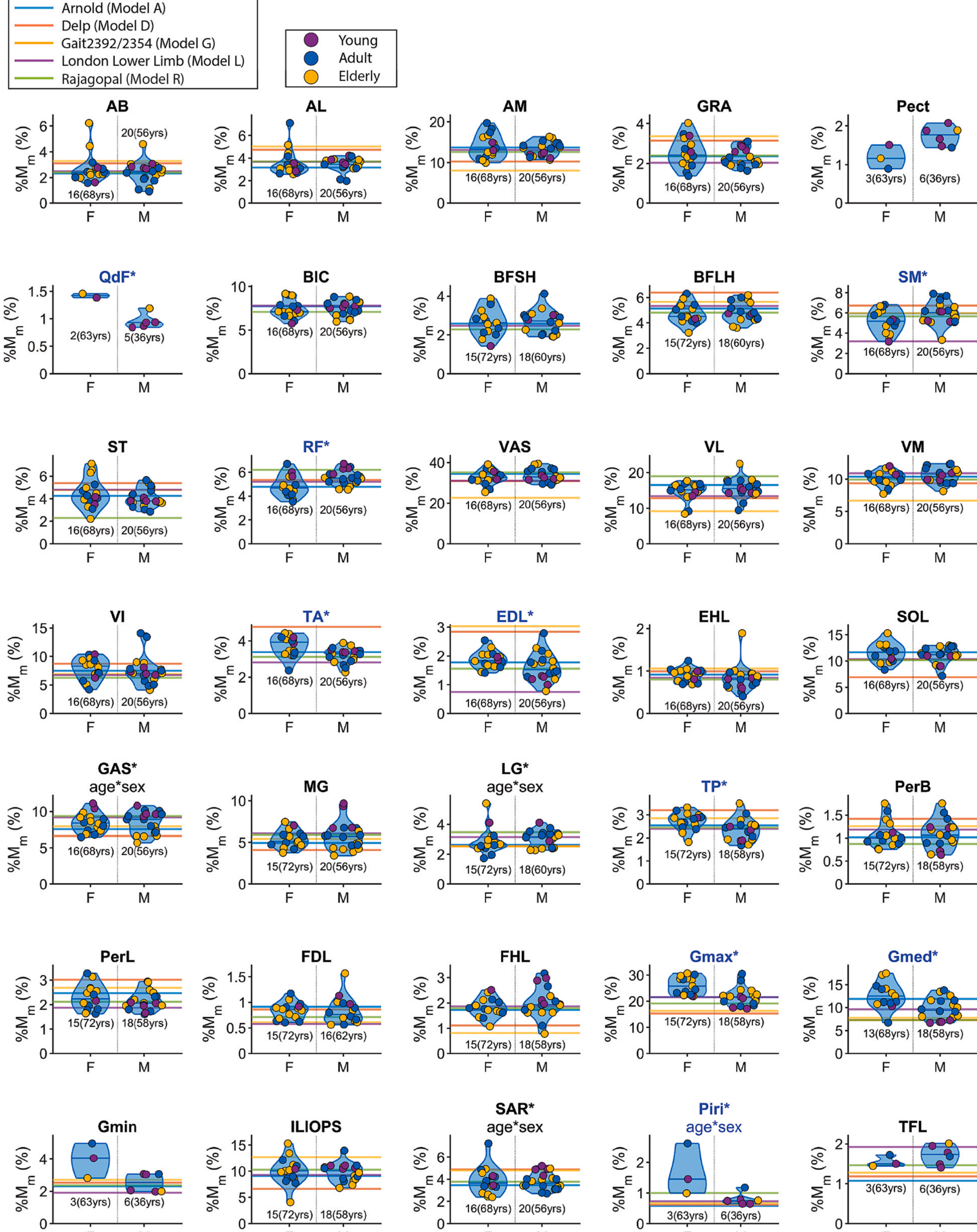

**Fig. 5.** Sex-related differences in relative muscle mass (%Mm) between females (F) and males (M). The colours in the dots indicate the age-groups (Young, Adult, Elderly). The title is coloured blue with an asterisk if a significant difference was found ($p < 0.05$). The horizontal lines indicate the muscle mass levels of the OpenSim opensource models. Some muscles were not incorporated in the musculoskeletal models (e.g. Pect, QdF). The number below the violin plots indicate the number of specimens (ex-vivo) included, this differs between muscles depending on the available data from the sources. In brackets is the average age, an asterisk there indicates that there is a significant difference in age between the groups. (For interpretation of the references to colour in this figure legend, the reader is referred to the web version of this article.)

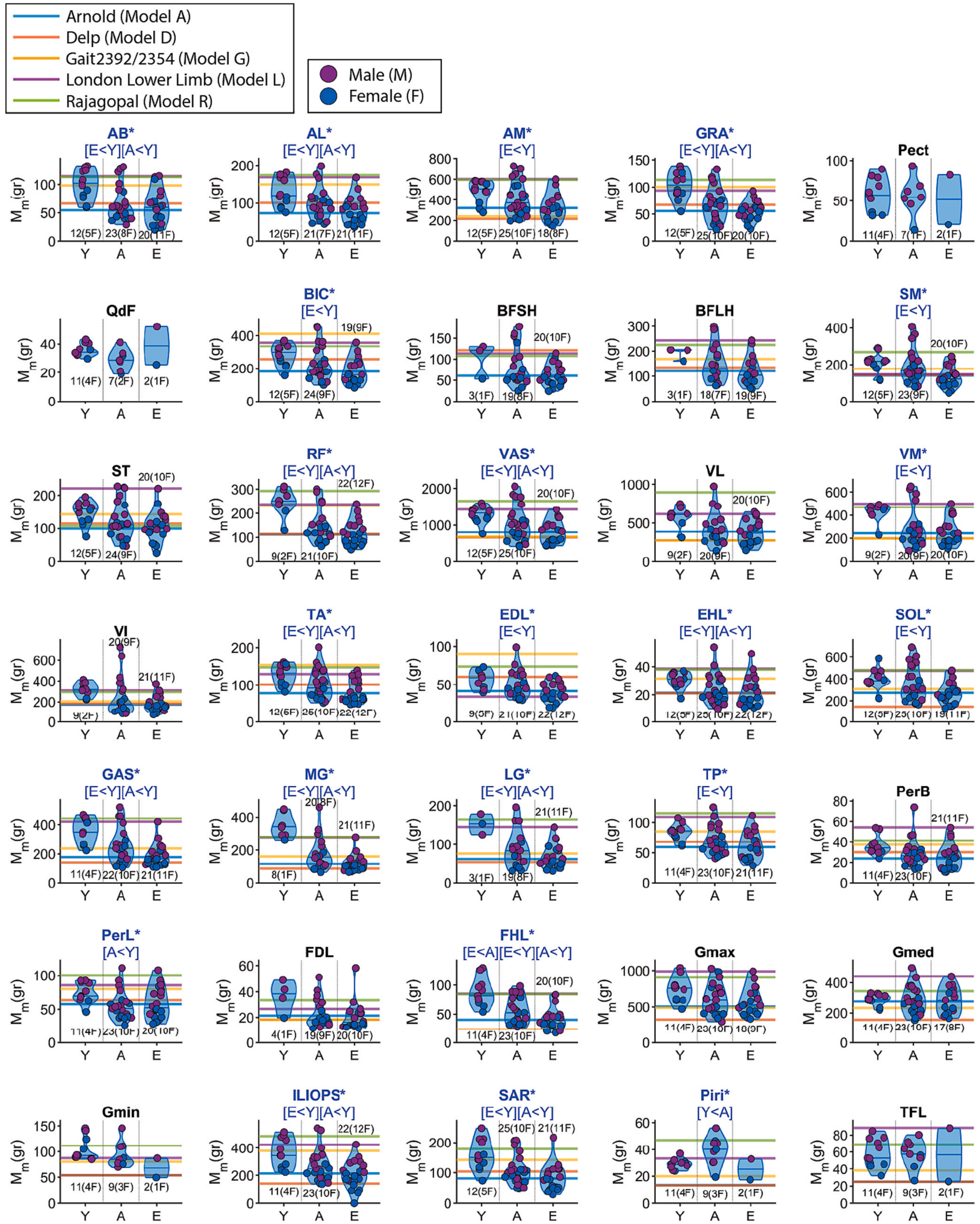

**Fig. 6.** Age-related differences in absolute muscles mass (Mm) between young (Y), adults (A), and elderly (E). The colours in the dots indicate the sex groups (Female, Male). The title is coloured blue with an asterisk if a significant difference was found ($p < 0.05$). The horizontal lines indicate the muscle mass levels of the OpenSim opensource models. Some muscles were not incorporated in the musculoskeletal models (e.g. Pect, QdF). The number below the violin plots indicate the number of specimens (ex-vivo) included within brackets the number of females in the group, this differs between muscles depending on the available data from the sources. (For interpretation of the references to colour in this figure legend, the reader is referred to the web version of this article.)

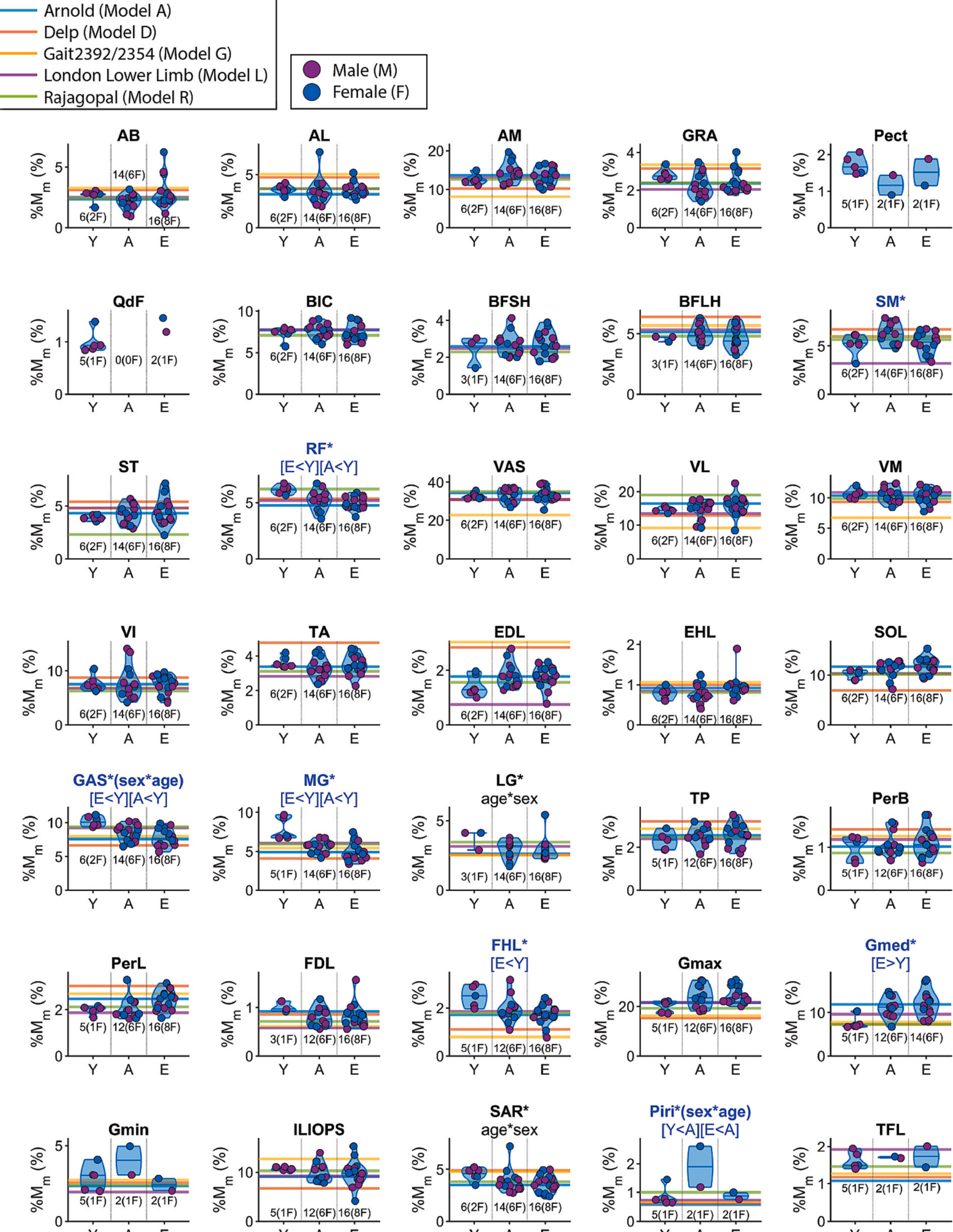

**Fig. 7.** Age-related differences in relative muscles mass between young (YA), adults (AD), and elderly (EL). The colours in the dots indicate the sex groups (Female, Male). The title is coloured blue if a significant difference was found ($p < 0.05$). The horizontal lines indicate the muscle mass levels of the OpenSim opensource models. Some muscles were not incorporated in the musculoskeletal models (e.g. Pect, QdF). The number below the violin plots indicate the number of specimens (ex-vivo) included, this differs between muscles depending on the available data from the sources. (For interpretation of the references to colour in this figure legend, the reader is referred to the web version of this article.)

### 4.3. Representation of age and sex in opensource musculoskeletal models

The age and sex differences in muscle strength distribution are generally not well represented in current opensource MSK models. We acknowledge that these models were developed at different times and often aim to build upon their predecessors (Londen Lower Limb being an improvement on Delp by using data from a single cadaver, Arnold being an improvement on Delp and London Lower Limb by using multiple cadavers, Rajagopal partly introduced data from young, healthy individuals (in-vivo)). Nonetheless, these successive iterations still do not offer targeted applicability to specific demographic groups, which limits their precision in most contexts (van der Kruk, 2025).

In particular, the Gait2392/2354 model inconsistently matched muscle mass distribution for either sex, and fluctuated between young and elderly muscle mass, making it unrepresentative of both groups. This inconsistency probably arises because for the Gait2392/2354 model additional strength scaling was applied to the Delp model to align with joint torque–angle relationships in living subjects (Anderson & Pandy, 1999; Carhart, 2000). Despite attempts to maintain a consistent scaling factor, different factors were ultimately used, especially for biarticular muscles.

For age representation, the Arnold model was most aligned with the elderly group mean for both Mm and %Mm. Regarding sex representation, it displayed Mm slightly above the female mean and below the male mean across all muscles. For muscles with sex-related differences in %Mm, the model represented females in RF, EDL, and Gmed, and males in TA, TP, and Gmax. Consequently, it does not adequately account for specific sex differences.

The Rajagopal and Londen Lower Limb models were representative of males for both Mm and %Mm, being the only two models demonstrating sex consistency. None of the models adequately represented females for %Mm. Rajagopal is also representative for Mm in the young and for %Mm in adults, making it the best choice for an average male non-elderly populations.

Although we expected the Londen Lower Limbs model's PCSA and Mm values to align with those typical of the elderly demographic, as it was based on a single elderly male, our results revealed that the Londen Lower Limb model exhibits higher PCSA and Mm across most muscles, aligning more closely with young male adults. The cadaver was selected for its higher muscularity compared to typical older cadavers, potentially accounting for this inconsistency. However, for %Mm, Londen Lower Limb is more representative of the elderly muscle mass distribution, particularly for muscles with significant age differences: Gmed, FHL, and GAS. This indicates that, while Mm values are elevated, the % Mm remains representative for the elderly age group. Translating PCSA into maximum isometric forces, Londen Lower Limb has a specific tension of 37 N/cm2, which is lower than that of other open-source models (60/61 N/cm2), resulting in lower maximum isometric forces compared to these models. Persad et al. (2024) suggest, based on an extensive review of human literature, that the appropriate value for human muscle-specific tension is 26.8 N/cm$^2$, indicating that specific tension values used in all models may still be subject to debate.

## 5. Limitations

- We assumed that %Mm is a reliable proxy for identifying age- and sex-related differences in %PCSA. However, the rules governing the scaling of optimal fiber length based on sex and size remain unclear. Son et al. (2024) demonstrated that while muscle mass scales with body mass, fiber length does not. This indicates that PCSA is unlikely to scale isometrically (to the 0.66 power) either. Due to the limited experimental data on these variations, we are unable to draw definitive conclusions at this time.
- We chose to scale the muscle mass as a percentage of total muscle mass, and not body mass. One could argue motion model predictions should be done based on proportion of total body mass, because a knowledge of this would allow to input appropriate muscle strength estimates for a given body mass, which might play an important role, especially in more dynamic predictions.
- We compiled a dataset of lower limb muscle mass from various sources (Supplementary Material A.6), with records dating back to 1884, when life expectancy was significantly lower (42 years for males and 44 years for females). Determining whether muscles experienced accelerated aging during that period is challenging. However, a comparison between the data from Theile (1884) and Son et al. (2024) reveals no apparent differences in adult muscle measurements. We do not expect that today's longer life expectancy correlates with better muscle health, as evidenced by the increasing prevalence of mobility issues in an aging society. Moreover, modern lifestyles are likely more sedentary. Thus, we have treated these groups as comparable in age while remaining aware of potential differences.
- A scarcity of PCSA data was identified, particularly considering the various methods used for calculation and the lack of data for specific groups, such as adults, young females, elderly males, children, and athletes. Muscle volume data are more prevalent, as this does not require determining optimal fiber length. A more extensive dataset could have been constructed by integrating muscle volume data with muscle architecture from different sources (Rajagopal et al., 2016). We chose not to include this method due to potential discrepancies among variables and sources.
- The statistical analysis was performed on data from dissection studies. Note that muscle volumes, and thus PCSA, appear to shrink post-mortem and may underestimate living PCSA (Friederich & Brand, 1990).
- The genetic background of participants is not always explicitly mentioned. Therefore, the current state-of-the-art may not be inclusive of all groups or may be biased toward a particular genetic background.
- This review specified four age groups. However, there is no consensus on the age boundaries per group, and one might argue that the adult group is too broad, losing individuality.

## 6. Conclusion

- Isometric scaling of maximum isometric forces in musculoskeletal models fails to account for significant age- and sex-related differences in muscle ratios (%Mm).
- Males have a higher proportion of muscle mass in the rectus femoris and semimembranosus compared to females, reflecting greater distribution toward these biarticular muscles in the lower limb. In contrast, females exhibit higher relative muscle mass in pelvic muscles (gluteus maximus and gluteus medius) and ankle muscles (tibialis anterior, tibialis posterior, and extensor digitorum longus).
- Older adults have a higher relative muscle mass in the gluteus medius than younger individuals, whereas young adults have greater relative muscle mass in the rectus femoris, (medial) gastrocnemius, and flexor hallucis longus.
- Current open-source musculoskeletal (MSK) models exhibit inconsistencies in representing age and sex in terms of absolute and relative muscle mass, with none accurately depicting female muscle mass distribution.
- There is a lack of sufficient data on the physiological cross-sectional area (PCSA), especially to determine %PCSA, which requires measurements on a complete leg to provide these data for the lower limb.

## CRediT authorship contribution statement

**R. Maarleveld:** Writing – review & editing, Writing – original draft,

Visualization, Validation, Investigation, Formal analysis, Data curation. **H.E.J. Veeger:** Writing – review & editing, Resources, Conceptualization. **F.C.T. van der Helm:** Writing – review & editing, Resources, Conceptualization. **J. Son:** Writing – review & editing, Resources, Data curation. **R.L. Lieber:** Writing – review & editing, Resources, Investigation. **E. van der Kruk:** Writing – review & editing, Visualization, Validation, Investigation, Formal analysis, Data curation, Resources.

## Declaration of competing interest

The authors declare that they have no known competing financial interests or personal relationships that could have appeared to influence the work reported in this paper.

## Acknowledgements

This study was funded by NWO-TTW VENI Grant 18145(2021). The funders had no role in study design, data collection and analysis, decision to publish, or preparation of the manuscript.

## Appendix A. Supplementary material

Supplementary data to this article can be found online at https://doi.org/10.1016/j.jbiomech.2025.112976.